\documentclass[pra,twocolumn,nofootinbib]{revtex4-1}
\usepackage{graphicx}
\usepackage{hyperref}
\usepackage[section]{placeins}
\usepackage{amsmath}
\usepackage{amsthm}
\usepackage{amssymb}
\usepackage{mathtools}
\usepackage{lmodern}
\usepackage{dsfont}
\usepackage{bbold}
\usepackage{bm}
\usepackage{paralist}
\usepackage{comment}
\usepackage{dutchcal}
\usepackage{color}
\newtheorem{theorem}{Theorem}

\newtheorem{prop}{Proposition}
\newcommand{\Tr}{\text{Tr}}

\newcommand{\bit}{\begin{itemize}}
\newcommand{\eit}{\end{itemize}\par\noindent}
\newcommand{\ben}{\begin{enumerate}}
\newcommand{\een}{\end{enumerate}\par\noindent}
\newcommand{\beq}{\begin{equation}}
\newcommand{\eeq}{\end{equation}\par\noindent}
\newcommand{\beqa}{\begin{eqnarray*}}
\newcommand{\eeqa}{\end{eqnarray*}\par\noindent}
\newcommand{\beqn}{\begin{eqnarray}}
\newcommand{\eeqn}{\end{eqnarray}\par\noindent}

\begin{document}

\title{From statistical proofs of the Kochen-Specker theorem to noise-robust noncontextuality inequalities}
\author{Ravi Kunjwal}
\email{rkunjwal@perimeterinstitute.ca}
\affiliation{Perimeter Institute for Theoretical Physics, 31 Caroline Street North, Waterloo, Ontario Canada N2L 2Y5}
\author{Robert W. Spekkens}
\email{rspekkens@perimeterinstitute.ca}
\affiliation{Perimeter Institute for Theoretical Physics, 31 Caroline Street North, Waterloo, Ontario Canada N2L 2Y5}

\date{\today}                                                

\begin{abstract}
The Kochen-Specker theorem rules out models of quantum theory wherein projective measurements are assigned outcomes deterministically and independently of context. This notion of noncontextuality is not applicable to experimental measurements because these are never free of noise and thus never truly projective.
For nonprojective measurements, therefore, one must drop the requirement that an outcome is assigned deterministically in the model and merely require that it is assigned a {\em distribution over outcomes} in a manner that is context-independent.  By demanding context-independence in the representation of preparations as well, one obtains a generalized principle of noncontextuality that also supports a quantum no-go theorem. 
Several recent works have shown how to derive inequalities on experimental data which, if violated, demonstrate the impossibility of finding a generalized-noncontextual model of this data. That is, these inequalities do not presume quantum theory and, in particular, they make sense without requiring an operational analogue of the quantum notion of projectiveness. 
We here describe a technique for deriving such inequalities starting from arbitrary proofs of the Kochen-Specker theorem.  It extends significantly previous techniques that worked only for logical proofs, which are based on sets of projective measurements that fail to admit of any deterministic noncontextual assignment, to the case of statistical proofs, which are based on sets of projective measurements that {\em do} admit of some deterministic noncontextual assignments, but not enough to explain the quantum statistics. 
\end{abstract}

\maketitle

\section{Introduction}
In quantum theory, a given sharp measurement (i.e., one associated to a projector-valued measure) can be compatible with each of two other sharp measurements that are incompatible with one another.  In this case, the latter pair of measurements define two distinct compatibility contexts for the first measurement.  The Kochen-Specker (KS) theorem \cite{KochenSpecker} rules out a particular kind of explanation of the operational predictions of quantum 
theory, namely, one wherein sharp measurements are assigned outcomes deterministically and independently of context.
 We will term this sort of explanation a {\em KS-noncontextual model} of the statistics.

A particular proof of the KS theorem is said to be {\em logical} if it appeals to the existence of a set of sharp quantum measurements that  admit of {\em no} deterministic noncontextual assignments.  (For the case where all measurement outcomes correspond to rank-1 projectors, i.e., projectors onto rays of Hilbert space, a deterministic noncontextual assignment is one that assigns, for every basis of orthogonal rays, the value 1 to precisely one such ray and the value 0 to the others, termed a {\em KS-colouring} of the rays.  In this case, a proof is said to be logical if it appeals to the existence of a set of rays in a Hilbert space that admit of {\em no KS-colourings} \cite{KochenSpecker, Peres, Mermin, Cabello18ray}.)
A proof is said to be {\em statistical} if it appeals to a set that {\em does} admit of some deterministic noncontextual assignments, but these assignments are insufficient to explain the statistics of the measurements on one or more quantum states. 
Explaining the statistics of the measurements here means recovering them as a convex mixture of deterministic noncontextual assignments to these measurements.

  As an aside, we note that in most of the literature on the subject (see, e.g., \cite{SICpaper}), a proof of the KS theorem is said to be {\em state-independent} if the proof works for an arbitrary choice of quantum state on which the measurements are implemented, 
 and it is said to be {\em state-dependent} if the proof appeals to a special quantum state. 
   In our terminology, {\em logical} proofs of the KS theorem are always state-independent.  The reason is that logical proofs by definition appeal to sets of sharp measurements that admit of {\em no} deterministic noncontextual assignments (for example, those that appeal to sets of rays that admit of no KS-colourings \cite{KochenSpecker, Cabello18ray}), and if the set of deterministic noncontextual assignments is empty, then the set of mixtures of deterministic noncontextual assignments is also empty, and we cannot explain measurement statistics in terms of such a mixture regardless of the quantum state.  {\em Statistical} proofs of the KS theorem, on the other hand, are those that appeal to sets of sharp measurements that admit of one or more deterministic noncontextual assignments (for example, sets of rays that admit of one or more KS-colourings).  These can be state-dependent, such as the proof in Ref.~\cite{KCBS}, or state-independent, such as the proof in Ref.~\cite{YuOh,SICpaper}.

In recent years, there has been much work deriving inequalities for operational statistics that follow directly from an 
assumption of noncontextuality, without assuming the validity of quantum theory. If these are violated experimentally, one can 
conclude that not just quantum theory, but any successor thereof must fail to admit of a noncontextual model.\footnote{The difference 
betwen a quantum no-go theorem for noncontextuality and a noncontextuality inequality is precisely analogous to the difference between
Bell's 1964 no-go theorem establishing the impossibility of a locally causal model of quantum theory \cite{Bell64} and the CHSH 
inequalities \cite{CHSH}, which are constraints on operational statistics that follow directly from the assumption of local causality, without
presuming the validity of quantum theory.}
One approach to deriving these inequalities 
is to seek inspiration from particular proofs of the KS theorem. 
Refs.~\cite{KunjSpek} and \cite{KrishnaSpekkensWolfe} do so for logical proofs.
Here, we consider statistical proofs. 

Section II reviews background material, in particular, the notions of operational theories, ontological models, measurement noncontextuality and Kochen-Specker (KS) noncontextuality. For pedagogical clarity, we present our result as a generalization of the results obtained previously for logical proofs \cite{KunjSpek}, but cast into a slightly different form. In Section III, therefore, we
 recast these earlier results.  Along the way, we review the notion of preparation noncontextuality and how it can be used to infer that certain measurements are assigned outcomes deterministically in the ontological model if certain operational correlations hold.  We also review how this inference allows for a proof of the failure of KS noncontextuality to be translated into a proof of the failure of preparation and measurement noncontextuality.  \color{black} Then, in Section IV, we generalize to the case of statistical proofs.  Our main result is presented in Theorem \ref{mainthm} and we give some concrete examples of its application. We close with a discussion in Section V.

Xu {\em et al.}~\cite{Xuetal} have previously obtained noise-robust noncontextuality inequalities for some statistical proofs of the KS theorem. Our approach in this article is distinct from that of Ref.~\cite{Xuetal} and allows a consideration of arbitrary statistical proofs of the KS theorem, rather than particular examples of it.  A discussion of the relation between Ref.~\cite{Xuetal} and this article is provided in Section V and Appendix \ref{commentxuetal}. \color{black}

\section{Preliminaries}
We conceptualize an experiment as a source followed by a measurement.  A source is a procedure that samples a random variable and implements a preparation on the system conditioned on the value of this variable.  The value of the variable is termed the {\em outcome} of the source.    The event of obtaining an outcome $s$ of a source $S$ will be termed a {\em source event}, denoted $[s|S]$. 
For each source event $[s|S]$, one can associate a preparation procedure on the system;
it is the one implemented by the source $S$ conditional on outcome $s$ occurring. The event of obtaining any outcome in some subset $V$ of all possible outcomes of a source $S$ will be denoted by $[V|S]$, and is associated to the preparation procedure wherein one implements $S$, coarse-grains over the outcomes in $V$,  and then conditions on obtaining this coarse-grained outcome. We denote the full set of outcomes by $\top$, so that the preparation procedure corresponding to implementing $S$ and {\em not} conditioning on obtaining any particular outcome is denoted $[\top| S]$.\footnote{Note that our notational conventions here align with those of Ref.~\cite{KrishnaSpekkensWolfe} rather than those of Ref.~\cite{KunjSpek}. For example, the preparation procedure $[\top|S]$ would be represented as $P_S^{\rm (ave)}$ in the notation of Ref.~\cite{KunjSpek}, where $S$ is the choice of source setting.} 
A measurement $M$ has outcome denoted by $m$, and the event of obtaining outcome $m$ of measurement $M$ is termed a {\em measurement event}, denoted $[m|M]$.   The measurement event $[V|M]$ for a subset $V$ of the outcomes is defined in the obvious manner, similarly to the case of sources. \color{black}

An {\em operational theory} specifies a rule for assigning a joint probability distribution $p(m,s|M,S)$, denoting the probability that in a prepare-and-measure experiment with source $S$ and measurement $M$, the source event $[s|S]$ occurs followed by the measurement event $[m|M]$.
For example, when the operational theory is quantum theory, the source event $[s|S]$ is represented by some density matrix, say $\rho_{[s|S]}$, the measurement event $[m|M]$ is represented by a positive operator, say $E_{[m|M]}$, and the rule for assigning the joint probability is $p(m,s|M,S)=p(s|S)\Tr(E_{[m|M]}\rho_{[s|S]})$, where $p(s|S)$ is the probability that the source $S$ yields outcome $s$. \color{black} Note that our analysis in this paper does not depend on the particular representation of preparations and measurements in an operational theory, nor on the particular probability rule associated with the theory; in this sense, we consider operational theories more general than quantum theory.

An {\em ontological model} of an operational theory posits that the causal influence of the source on the measurement is mediated by the ontic state, $\lambda$, of the system (a point in the underlying ontic state space $\Lambda$, which for our purposes can be taken to be discrete).
For a source $S$ with outcome $s$, the ontological model associates a 
conditional probability $\mu(\lambda,s|S)$ such that 
$\sum_s \sum_{\lambda\in \Lambda} \mu(\lambda,s|S) =1$. Here $\mu(\lambda,s|S)=\mu(\lambda|s,S)p(s|S)$.
For a measurement $M$ with outcome $m$, the ontological model associates
a conditional probability $\xi(m|M,\lambda)$ such that $\sum_{m} \xi(m|M,\lambda)=1$ for all $\lambda \in \Lambda$.  Finally, the ontological model must reproduce the statistical predictions of the operational theory, 
\begin{equation}\label{eq:empiricaladequacy}
{\rm pr}(m,s|M,S) = \sum_{\lambda\in \Lambda}\xi(m|M,\lambda)\mu(\lambda,s|S).
\end{equation}

 In the case of quantum theory, two measurement procedures differ {\em only} by context if and only if they yield the same statistics {\em for all quantum states}. In this case, they are represented by the {\em same} positive operator-valued measure (POVM).\footnote{The type of measurement that is considered in most discussions of the Kochen-Specker theorem is a projective measurement (which we here refer to as a {\em sharp} measurement). Such measurements are a special class of POVMs wherein the positive operators are all projectors.  Note that these are the only measurements in quantum theory that can be represented by a single Hermitian operator, namely, the one whose spectral projectors are the elements of the projector-valued measure.}
Equivalently, two measurement {\em events} differ only by context   if and only if they are assigned the same probability by all quantum states. In this case, they are represented by the same positive operator less than identity (or the same projector in the case of a sharp measurement).

By analogy, in an arbitrary operational theory, two  measurement events, $[m|M]$ and $[m'|M']$, differ only by context if and only if for every preparation procedure, the probability of $[m|M]$ is the same as that of $[m'|M']$.
The condition can be formalized as:\footnote{By a Bayesian inversion, this condition is equivalent to  $\forall [s|S]: {\rm pr}(m,s|M,S)= {\rm pr}(m',s|M',S)$, a form that makes more apparent the close analogy with the operational equivalence relation among source events which we define further on.} 
\begin{align}\label{OEmmts}
\forall [s|S]: {\rm pr}(m|M,s,S)= {\rm pr}(m'|M',s,S).
\end{align}
When measurement events $[m|M]$ and $[m'|M']$ differ only by context, they are said to be {\em operationally equivalent}, denoted $[m|M] \simeq [m'|M']$.

In Ref.~\cite{Spe05}, a {\em measurement noncontextual} ontological model was defined to be one wherein operationally equivalent measurement events are represented by equivalent response functions:
\begin{align}\label{MNC}
[m|M] \simeq [m'|M'] \implies  \xi(m|M,\lambda) = \xi(m'|M',\lambda),\forall\lambda\in\Lambda,
 \end{align}
where one allows the measurements to respond {\em in}deterministically to $\lambda$:
$ \xi(m|M,\lambda) \in [0,1]$.  
On the other hand, many have proposed to generalize the notion of KS-noncontextuality from quantum theory to arbitrary operational theories in a different manner, namely, by assuming
Eq.~\eqref{MNC} but with measurements responding deterministically, that is,
\begin{align}\label{OD}
\xi(m|M,\lambda) \in \{ 0,1\}.
 \end{align}
We term the latter proposal {\em KS-noncontextuality}:
\begin{align}\label{KSNC}
&\textrm{KS-noncontextuality} \nonumber\\
&= \textrm{Measurement noncontextuality}\;(\textrm{Eq.}~\eqref{MNC})\nonumber\\
&+ \textrm{Outcome determinism}\; (\textrm{Eq.}~\eqref{OD})\quad\forall \lambda\in\Lambda.
\end{align}

In the following, we let $\frak{M}$ denote a set of measurement procedures whose operational features include the compatibility relations and operational equivalences that their quantum counterparts satisfy in a proof of the KS theorem.

To operationalize the KS theorem, the notion of compatibility must also be generalized to an arbitrary operational theory.
We follow the proposal of Ref.~\cite{LSW}: measurements $M_1$ and $M_2$ are deemed compatible if there is a third measurement with an outcome set that is the Cartesian product of the two outcome sets such that one simulates $M_1$ and $M_2$ by marginalization (i.e., for which the marginalized versions are operationally equivalent to  $M_1$ and $M_2$).

\section{From logical proofs of the KS theorem to operational criteria for universal noncontextuality}

It follows from the above that the proof schema which generalizes a logical proof of the KS theorem from quantum theory to an arbitrary operational theory is of the following form:

\begin{prop}[No-go for KS noncontextuality from logical proof]\label{ksuncolnogo}\hfill\newline
Measurement noncontextuality (Eq.~\eqref{MNC})\\
$+$ Outcome determinism (Eq.~\eqref{OD}) $\forall \lambda\in\Lambda, \forall M \in \frak{M}$\\
$+$ Operational equivalences in the set $\frak{M}$ (proof dependent)\\
$\implies$ Contradiction.
\end{prop}

In the quantum case, this contradiction can be inferred from KS-uncolourability; Kochen and Specker's original proof of the KS theorem is an example~\cite{KochenSpecker}.\\

Note that in the face of this contradiction, one can always salvage the spirit of noncontextuality (measurement noncontextuality) 
simply by abandoning outcome determinism.
By contrast, this is not a way out of Bell's theorem because the notion of local causality does not presume outcome determinism.

For these reasons, it was argued in Ref.~\cite{Spe05} that one should {\em drop} the assumption of outcome determinism that is part of 
KS-noncontextuality and simply assume measurement noncontextuality.
Such a move blocks the derivation of the contradiction in Proposition \ref{ksuncolnogo}. It might appear, therefore,
 that there is in fact {\em no} conflict between quantum theory and the spirit of noncontextuality if one excises the notion of outcome determinism from the latter.  However, it turns out that the property of outcome determinism can be {\em inferred} for certain measurements by applying a notion of noncontextuality to preparations~\cite{Spe05}, as we now explain.

Two source events, $[s|S]$ and $[s'|S']$, are operationally equivalent, denoted $[s|S] \simeq [s'|S']$, if for every measurement event $[m|M]$, the joint probability of obtaining $[s|S]$ and $[m|M]$ is the same as that of obtaining $[s'|S']$ and $[m|M]$,
\begin{align}\label{OEpreps}
\forall [m|M]: p(m,s|M,S)=p(m,s'|M,S').
\end{align}

The assumption of preparation noncontextuality  requires that operationally equivalent source events should be represented equivalently in the ontological model:\footnote{
Note that Eq.~\eqref{OEpreps} is analogous to Eq.~\eqref{OEmmts} and Eq.~\eqref{PNC} is analogous to Eq.~\eqref{MNC}.
} 
\begin{align}\label{PNC}
[s|S] \simeq [s'|S'] \implies  \mu(\lambda,s|S) =  \mu( \lambda,s'|S'),\quad\forall\lambda\in\Lambda.
 \end{align}

 It was argued in Ref.~\cite{Spe05} that whatever reasons can be given in support of measurement noncontextuality, these are also reasons to believe in preparation noncontextuality and therefore that the only reasonable assumption to make is noncontextuality for all experimental procedures, termed {\em universal noncontextuality}.  It is the assumption we make here.
 
Ref.~\cite{Spe14} showed that for quantum theory, preparation noncontextuality implies that measurements should be assigned outcomes deterministically if and only if they are projective. Ref.~\cite{KunjSpek} generalized the logic to all operational theories by focusing not just on the 
set $\frak{M}$ of measurements, but on a corresponding set $\frak{S}$ of sources as well.
Outcome determinism is justified when the following operational criteria are satisfied:

(i) For each equivalence class of measurements,
$\mathcal{M}_i\in \frak{M}$, there must exist an equivalence class of sources, $\mathcal{S}_i \in \frak{S}$, such that the outcomes of $\mathcal{M}_i$ and $\mathcal{S}_i$, denoted $\mathcal{m}_i$ and $\mathcal{s}_i$ respectively, are perfectly correlated. This implies that the average correlation over the pairings $\{(\mathcal{M}_i, \mathcal{S}_i): i\in\{ 1,\ldots,n\}\}$,
\begin{align}\label{defnCorr}
{\rm Corr} \equiv \frac{1}{n}\sum_{i=1}^{n} \sum_{\mathcal{m}_i,\mathcal{s}_i}\delta_{\mathcal{m}_i,\mathcal{s}_i } \rm{pr}( \mathcal{m}_i, \mathcal{s}_i |\mathcal{M}_i, \mathcal{S}_i),
\end{align}
satisfies
\begin{align}\label{perfCorr}
\rm{Corr} = 1.
\end{align}

(ii)  
The sources must obey the following operational equivalence relations:
\begin{align}\label{opeqSs}
\forall i,i': [\top|\mathcal{S}_i]\simeq [\top |\mathcal{S}_{i'}],
\end{align}
where, as stipulated earlier, $[\top|\mathcal{S}]$ denotes the event corresponding to implementing $\mathcal{S}$ and not conditioning on its outcome.

Note that if the source event $[s|S]$ is represented in the ontological model by $\mu(\lambda,s|S)$, then $[\top|S]$ is represented by 
\begin{align}\label{marginalization}
\mu(\lambda|S) \equiv  \sum_{s} \mu(\lambda,s|S).
\end{align}
The inference established in Ref.~\cite{KunjSpek} can then be expressed as follows:
\begin{prop}[Justifying outcome determinism]\label{inference}\hfill\newline
Preparation noncontextuality (Eq.~\eqref{PNC})\\
+Operational equivalences in the set $\frak{S}$ (Eq.~\eqref{opeqSs})\\
+Perfect Correlation between outcomes of $\mathcal{S}_i$ and $\mathcal{M}_i$ 
for all $i$ (Eq.~\eqref{perfCorr})\\
$\implies$ Outcome determinism $\forall\lambda \in \cup_{\mathcal{S}\in \frak{S}} \textrm{supp} (\mu(\cdot|\mathcal{S}))$ (Eqs.~\eqref{ODsuppmu},\eqref{supportsources}) $\forall M \in \frak{M}$.
\end{prop}

The proof is as follows. 
From Eqs.~\eqref{PNC} and \eqref{opeqSs}  we conclude that
\begin{align}\label{commonsupports}
\forall i,i' : \mu(\lambda |\mathcal{S}_{i}) &=\mu(\lambda |\mathcal{S}_{i'})\nonumber\\
&\equiv \nu(\lambda).
\end{align} 
By Bayesian inversion, $  \mu(\mathcal{s}|\lambda,\mathcal{S}) = \mu(\lambda,\mathcal{s}|\mathcal{S})/ \nu(\lambda)$.
Substituting this into Eq.~\eqref{eq:empiricaladequacy}, we have
\begin{align}\label{probsInOM}
{\rm pr}(\mathcal{m}_i,\mathcal{s}_i | \mathcal{M}_i,\mathcal{S}_i ) 
&= \sum_{\lambda\in\Lambda} \xi(\mathcal{m}_i|\mathcal{M}_i , \lambda) \mu(\mathcal{s}_i|\lambda ,\mathcal{S}_i) \nu(\lambda).
\end{align}
Given this expression, the only way to explain the perfect correlation of Eq.~\eqref{perfCorr}, then, is if the measurements respond deterministically for all ontic states in the support of $\nu$, that is,
\begin{align}\label{ODsuppmu}
\forall \lambda \in \textrm{supp}(\nu): \xi(\mathcal{m}_i|\mathcal{M}_i , \lambda) \in \{0,1\} .
\end{align}
where $\textrm{supp}(\nu) \equiv \{ \lambda \in \Lambda: \nu(\lambda)>0\}$.
But given Eq.~\eqref{commonsupports},
\begin{align}\label{supportsources}
\textrm{supp}(\nu) =  \textrm{supp} (\mu(\cdot|\mathcal{S})) \;\forall \mathcal{S}\in \frak{S},
\end{align}
which concludes the proof.

To obtain a contradiction in the no-go theorem of Proposition \ref{ksuncolnogo}, it is sufficient to assume outcome determinism just for the 
ontic states in the union of the ontic supports of the distributions representing the sources in $\frak{S}$, even though this might be a subset of
the full set of ontic states. This is because any ontic state outside this subset
 does not affect 
the operational statistics of any experiment involving measurements on sources in $\frak{S}$.
Hence, the premiss of outcome determinism $\forall \lambda \in \Lambda$ in the no-go of Proposition \ref{ksuncolnogo} can be replaced by the same premiss $\forall\lambda \in \cup_{\mathcal{S}\in \frak{S}} \textrm{supp} (\mu(\cdot|\mathcal{S}))$ and thus by the 
antecedent of Proposition \ref{inference}.

The no-go that one obtains by combining Proposition \ref{inference} with Proposition \ref{ksuncolnogo} is a no-go for universal noncontextuality based on a logical proof of the KS theorem.

\begin{prop}[No-go for universal noncontextuality from logical proof]\label{logicalnogouniversalNC}\hfill\newline
Universal noncontextuality (Eq.~\eqref{MNC}, Eq.~\eqref{PNC})\\
+Operational equivalences in the set $\frak{S}$ (Eq.~\eqref{opeqSs})\\
+Operational equivalences in the set $\frak{M}$ (proof dependent)\\
+Perfect Correlation between outcomes of $\mathcal{S}_i$ and $\mathcal{M}_i$
for all $i$ (Eq.~\eqref{perfCorr})\\
$\implies$ Contradiction.
\end{prop}

As noted in Ref.~\cite{KunjSpek},  it implies that any operational theory that {\em does} admit of a universally noncontextual model while exhibiting the appropriate operational features of 
$\frak{M}$ and $\frak{S}$  must exhibit {\em imperfect} correlations for the pairings $\{(\mathcal{S}_i ,\mathcal{M}_i) \}$, that is, it 
must satisfy ${\rm Corr}<1$.  

 The precise amount by which ${\rm Corr}$ is bounded away from 1 is determined as follows. 
Substituting Eq.~\eqref{probsInOM} into the definition of ${\rm Corr}$ (Eq.~\eqref{defnCorr}),
 we obtain
\begin{align}\label{ExpForCorr}
\textrm{Corr} = \sum_{\lambda} \textrm{Corr}(\lambda) \nu(\lambda),
\end{align}
where 
\begin{align}\label{ExpForCorrlambda}
\textrm{Corr}(\lambda) &\equiv \frac{1}{n}\sum_{i=1}^{n} \sum_{\mathcal{m}_i,\mathcal{s}_i}\delta_{\mathcal{m}_i,\mathcal{s}_i } \xi(\mathcal{m}_i|\mathcal{M}_i,\lambda)\mu(\mathcal{s}_i | \mathcal{S}_i, \lambda).
\end{align}
For a given choice of $\lambda$ corresponding to a noncontextual assignment to the $\{ \mathcal{M}_i\}_i$,
$\textrm{Corr}(\lambda)$ is maximized by taking $\mu(\mathcal{s}_i | \mathcal{S}_i, \lambda)=1$ for $\mathcal{s}_i = \mathcal{m}_i^{\rm max}$, where $\mathcal{m}_i^{\rm max}$ is any value of $\mathcal{m}_i$ such that $\max_{\mathcal{m}_i} \xi(\mathcal{m}_i|\mathcal{M}_i,\lambda)=\xi(\mathcal{m}_i^{\rm max}|\mathcal{M}_i,\lambda)$. We then have $\textrm{Corr}(\lambda) = \frac{1}{n}\sum_{i=1}^{n} \max_{m_i} \xi(\mathcal{m}_i|\mathcal{M}_i,\lambda)$.  Because every noncontextual assignment is indeterministic for some $\mathcal{M}_i$, $\textrm{Corr}(\lambda)$ is bounded away from 1 for all $\lambda$.  Letting ${\rm Corr}_{\rm ind}$ denote the maximum value of $\frac{1}{n}\sum_{i=1}^{n} \max_{m_i} \xi(\mathcal{m}_i|\mathcal{M}_i,\lambda)$ in a variation over $\lambda$ that correspond to indeterministic noncontextual assignments, we have 
\begin{align}\label{NCIlogical}
\textrm{Corr} \le {\rm Corr}_{\rm ind}.
\end{align}

The qualifier that $\lambda$ correspond to indeterministic noncontextual assignments in the variation over $\lambda$ that defines ${\rm Corr}_{\rm ind}$ may seem unnecessary at this stage since {\em all} $\lambda$ in a logical proof of the KS theorem correspond to such assignments. However, we will soon consider statistical proofs of the KS theorem, where there exist $\lambda$ that correspond to deterministic noncontextual assignments and 
where the qualifier that ${\rm Corr}_{\rm ind}$ is computed by varying over $\lambda$ that correspond to indeterministic noncontextual assignments becomes necessary.

The compatibility and operational equivalence relations on the $\{ \mathcal{M}_i \}_i$, combined with the assumption of measurement noncontextuality, define linear constraints on the $n$-tuple of response functions $\{ \xi(\mathcal{m}_i|\mathcal{M}_i,\lambda) \}_i$. These linear constraints describe the facets of a polytope, termed the noncontextual measurement-assignment polytope.  One can obtain the vertices of this polytope from its facets using convex hull algorithms.  One determines ${\rm Corr}_{\rm ind}$ by determining the maximum value of $\frac{1}{n}\sum_{i=1}^{n} \max_{m_i} \xi(\mathcal{m}_i|\mathcal{M}_i,\lambda)$ in a variation over the 
(indeterministic) 
vertices.   The noncontextuality inequality one obtains for a given logical proof of the KS theorem, therefore, is simply the inequality one obtains by substituting the determined value of ${\rm Corr}_{\rm ind}$ into Eq.~\eqref{NCIlogical}.  Refs.~\cite{KunjSpek,KrishnaSpekkensWolfe} provide examples of how to derive such inequalities for specific logical proofs of the KS theorem.   

\section{From statistical proofs of the KS theorem to operational criteria for universal noncontextuality}

We can now turn to the question of how to obtain 
operational criteria for the failure of
universal noncontextuality from statistical, rather than logical, proofs of the KS theorem.
In the operationalized version of such proofs (i.e., one that makes no reference to the quantum formalism), the contradiction is achieved by noting that there is a special source event, that is,
the event of obtaining a special outcome $\mathcal{s}_*=0$ of a special source $\mathcal{S}_*$, denoted $[\mathcal{s}_*=0|\mathcal{S}_*]$, such that 
the measurement statistics one obtains for the special preparation associated to this event are inconsistent with KS-noncontextuality.
 
We index the compatible subsets of $\frak{M}$ by $\alpha$ and denote the equivalence class of measurements that jointly simulates the elements of this subset by
$\mathcal{M}^{(\alpha)}$, with the vector of outcomes denoted by  $\vec{\mathcal{m}}^{(\alpha)}$.

Let $R$ be the value of a particular linear function $F$ of the operational statistics for the compatible subsets of measurements when these are implemented on the special preparation, i.e.,
\begin{align}\label{defnR}
R = F( \{ \textrm{pr}( \vec{\mathcal{m}}^{(\alpha)} | \mathcal{M}^{(\alpha)}, \mathcal{s}_{*}=0, \mathcal{S}_{*} ) \}_{\alpha} ).
\end{align}
We define $R_{\rm det}$ to be the largest value of $R$ consistent with {\em deterministic} noncontextual measurement assignments, and $R_{\rm ind}$ to be the largest value of $R$ consistent with {\em indeterministic} noncontextual measurement assignments.
 For every statistical proof of the KS theorem, it is possible to construct a function $F$ such that
\begin{align}\label{Rbound}
R_{\rm ind} \ge R > R_{\rm det}.
\end{align}
(An example is given in Eq.~\eqref{defnRncycle}.)  

We define 
\begin{align}\label{pstar}
p_* \equiv {\rm pr}(\mathcal{s}_*=0|\mathcal{S}_*)
\end{align}
 to be the probability of the source $\mathcal{S}_*$ yielding the outcome $\mathcal{s}_*=0$.  The assumption that the special preparation sometimes occurs can be formalized as
\begin{align}\label{nonzeroprob}
p_* > 0.
\end{align}

Recalling ~\eqref{KSNC}, it follows that the proof schema for a no-go theorem for KS noncontextuality based on a statistical proof of the KS theorem is as follows:
\begin{prop}[No-go for KS noncontextuality from statistical proof]\label{kscolnogo}\hfill\newline
Measurement noncontextuality (Eq.~\eqref{MNC})\\
+Outcome determinism (Eq.~\eqref{OD}) $\forall \lambda\in \Lambda, \forall M \in \frak{M}$\\
+Operational equivalences in the set $\frak{M}$ (proof-dependent)\\
+Features of correlations among compatible subsets of $\frak{M}$ for the special preparation (Eq.~\eqref{Rbound})\\
+Nonzero probability of the special preparation (Eq.~\eqref{nonzeroprob})\\
$\implies$ Contradiction.
\end{prop}

A simple example of such a no-go theorem is based on the $n$-cycle scenario for odd $n$~\cite{KCBS, LSW, allncycles, CSW}. Here, there are $n$ equivalence classes of binary-outcome measurements, $\frak{M} \equiv \{ \mathcal{M}_i \}_{i=1}^n$,
where adjacent pairs are compatible, so that there are $n$ compatible subsets, which we can index by $\{(1,2),(2,3), \dots, (n-1,n), (n,1) \}$.  Let $\mathcal{M}^{(i, i \oplus 1)}$ denote the equivalence class of measurements which jointly simulates $\mathcal{M}_{i}$ and $\mathcal{M}_{i \oplus 1}$
  (here, $\oplus$ denotes sum modulo $n$).
Let $M^{(i,i\oplus 1)}$  denote a procedure in the equivalence class $\mathcal{M}^{(i,i \oplus 1)}$, and let $M_{i(i\oplus 1)}$ be the procedure one obtains by implementing $M^{(i,i\oplus 1)}$ and marginalizing over the outcome of $\mathcal{M}_{i \oplus 1}$. Note that by this definition, $M_{i(i\oplus 1)}$ is in the equivalence class  $\mathcal{M}_i$.
Define $M_{i(i\ominus 1)}$ similarly.  
 The difference between $M_{i(i\oplus 1)}$  and $M_{i(i\ominus 1)}$ is merely a difference of context, corresponding to the neighbour with which $\mathcal{M}_i$ is jointly implemented. 
The relevant operational equivalence relations (implicit in the definition of the equivalence classes) are therefore
$\forall i : M_{i(i\oplus 1)} \simeq M_{i(i\ominus 1)}$.  An assignment of outcomes  (deterministic or indeterministic) to these measurements in the ontological model  is noncontextual if it is independent of this choice.  

Clearly, if $n$ is odd, then not all adjacent pairs of measurements can have anticorrelated outcomes if the assignment is deterministic. 
At most, this can occur for $n-1$ out of the $n$ pairs.
Consequently, if one defines $R$ to be the probability of seeing anticorrelated 
outcomes when jointly implementing an adjacent pair of measurements ($ \mathcal{M}_i$ and $\mathcal{M}_{i\oplus 1}$) on the special preparation,
 averaged over all such pairs, i.e.,
\begin{align}\label{defnRncycle}
R &\equiv \frac{1}{n} \sum_{i=1}^{n} {\rm pr}( \mathcal{m}_i \ne \mathcal{m}_{i\oplus 1}| \mathcal{M}^{(i,i\oplus 1)}, \mathcal{s}_*=0, \mathcal{S}_* ),
\end{align}
then the largest value achievable by deterministic noncontextual assignments is $R_{\rm det}= \frac{n-1}{n}$, and KS-noncontextuality implies $R \le \frac{n-1}{n}$.   It follows that in any operational theory that predicts $p_* >0$ and
$R> \frac{n-1}{n}$, we obtain a contradiction with KS-noncontextuality.  

\begin{figure}
 \includegraphics[scale=0.3]{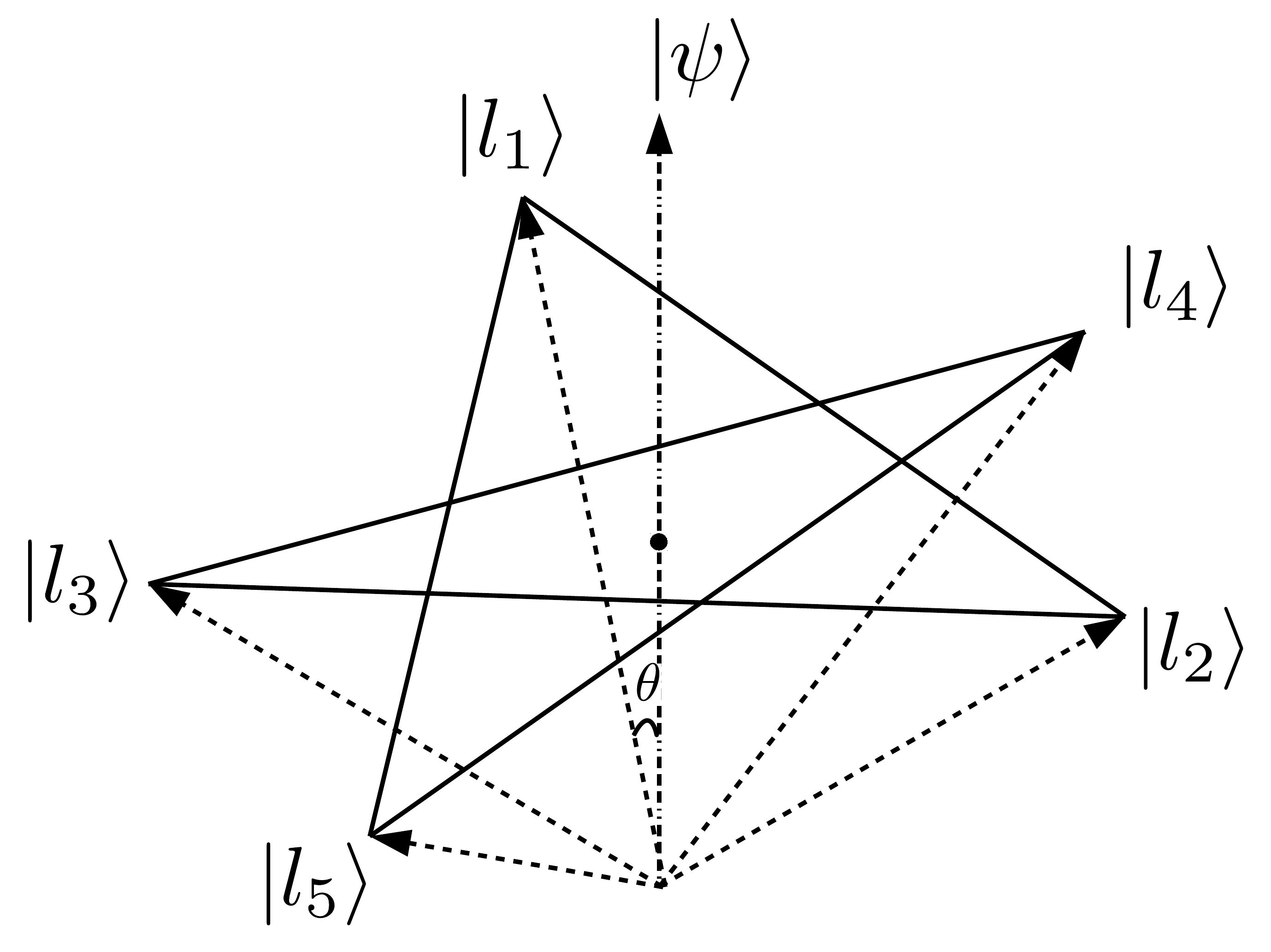}
 \caption{Quantum construction for a statistical proof of the KS theorem based on the 5-cycle~\cite{KCBS}.}
 \label{kcbsconfig}
 \end{figure}

As is well known, this is the case for quantum theory.
The first instance of such a proof, due to Klyachko, Can, Binicioglu and Shumovsky (KCBS)~\cite{KCBS}, was for the 5-cycle.
It showed that the compatibility relations required to hold among the $\{ \mathcal{M}_i \}_{i=1}^5$ can be achieved with sharp quantum measurements on a qutrit if $\mathcal{M}_i$ corresponds
to the projector-valued measure $\{ |l_i\rangle\langle l_i| , \mathbb{1} - |l_i\rangle\langle l_i| \}$, where 
$|l_i\rangle=(\sin\theta \cos\phi_i,\sin\theta \sin\phi_i,\cos\theta)$, $\phi_i=\frac{4\pi i}{5}$, and $\cos\theta=\frac{1}{\sqrt[4]{5}}$.
These are depicted as 3-dimensional vectors in Fig.~\ref{kcbsconfig}.
The special preparation event $[\mathcal{s}_*=0|\mathcal{S}_*] $ corresponds to the quantum state $|\psi\rangle=(0,0,1)$, also depicted in Fig.~\ref{kcbsconfig}.   Consequently, by letting $\mathcal{S}_*$ be a quantum source that prepares $|\psi\rangle$ with nonzero probability, we ensure that $p_*>0$.
The average anticorrelation for $|\psi\rangle$ is found to be $R=\frac{2}{\sqrt{5}}\approx 0.89442$, contradicting the prediction of KS-noncontextuality that 
$R \le \frac{4}{5}$. 
All of this generalizes to arbitrary odd $n\geq5$ \cite{LSW,allncycles,CSW}: the $|l_i\rangle$ have the same form with
$i\in\{1,2,\dots,n\}$, $\phi_i=\frac{n-1}{n}\pi i$, and $\cos^2\theta=\cos(\pi/n)/(1+\cos(\pi/n))$, with $|\psi\rangle$ as before,
 so that $p_*>0$ and $R=\frac{2\cos(\frac{\pi}{n})}{1+\cos(\frac{\pi}{n})}>\frac{n-1}{n}$.

To convert a statistical proof of the KS theorem 
 into a proof of the failure of universal noncontextuality, we proceed analogously to the conversion procedure for logical proofs. 
 We note that in the no-go of Proposition \ref{kscolnogo}, one can continue to derive a contradiction if one assumes outcome determinism just for the ontic states in the union of the supports of the distributions for the sources in the set $\frak{S}$ (redefined to include the special source $\mathcal{S}_*$), as long as one assumes that the operational features of the set $\frak{S}$ now include the fact that $[\top|\mathcal{S}_*]$ is operationally equivalent with all other marginalized sources,
\begin{align}\label{opeqSs2}
\forall i,i' : [\top|\mathcal{S}_i]\simeq [\top |\mathcal{S}_i']  \simeq [\top|\mathcal{S}_*].
\end{align}

Therefore, among the premisses of Proposition \ref{kscolnogo}, we can replace the assumption of outcome determinism with the antecedent of the
inference of Proposition \ref{inference} (with $\frak{S}$ now including $\mathcal{S}_*$ and Eq.~\eqref{opeqSs} replaced by Eq.~\eqref{opeqSs2}), to obtain a no-go theorem for universal noncontextuality:
\begin{prop}[No-go for universal noncontextuality from statistical proof]\label{nogouniversalNC}\hfill\newline
Universal noncontextuality (Eq.~\eqref{MNC}, Eq.~\eqref{PNC})\\
+Operational equivalences in the set $\frak{S}$ (Eq.~\eqref{opeqSs2})\\
+Operational equivalences in the set $\frak{M}$ (proof dependent)\\
+Perfect Correlation between outcomes of $\mathcal{S}_i$ and $\mathcal{M}_i$ for all $i$ (Eq.~\eqref{perfCorr})\\
+Features of correlations among compatible subsets of $\frak{M}$ for the special preparation (\textrm{Eq.}~\eqref{Rbound})\\
+Nonzero probability of the special preparation (Eq.~\eqref{nonzeroprob})\\ 
$\implies$ Contradiction.
\end{prop}

 It is useful to see how this proof schema yields a no-go for universal noncontextuality in quantum theory when one particularizes to the $n$-cycle scenario.  Let $\{ \mathcal{M}_i \}_i$ be the set of projective binary-outcome measurements on a qutrit specified earlier.
Let $\mathcal{S}_i$ be the quantum source that prepares $ |l_i\rangle\langle l_i|$ with probability $\frac{1}{3}$ and 
$\frac{\mathbb{1}-|l_i\rangle\langle l_i|}{2}$ with probability $\frac{2}{3}$. Similarly, let $\mathcal{S}_*$ be the quantum source that prepares $ |\psi\rangle\langle \psi|$  with probability $\frac{1}{3}$ (corresponding to outcome $\mathcal{s}_*=0$) and 
$\frac{\mathbb{1}-|\psi\rangle\langle \psi|}{2}$ with probability $\frac{2}{3}$. 
Clearly, we have the operational equivalences of \eqref{opeqSs2} by virtue of the fact that all of these ensembles average to $\frac{1}{3}\mathbb{1}$. Furthermore, we have ${\rm Corr}=1$.  
Thus all of the antecedents of the inference of 
Proposition \ref{nogouniversalNC} are satisfied, and we have a proof of the failure of universal noncontextuality in quantum theory. 
 
With the proof schema of Proposition \ref{nogouniversalNC} in hand, we can finally turn to the question of how to derive a noncontextuality inequality from statistical proofs of the KS theorem.

It suffices to note that Proposition \ref{nogouniversalNC} implies that any operational theory that {\em does} admit of a universally noncontextual 
model while exhibiting the specified operational features of the sets $\frak{M}$ and $\frak{S}$ cannot satisfy 
 Eqs.~\eqref{perfCorr}, \eqref{Rbound} and \eqref{nonzeroprob}, that is, it cannot satisfy $\textrm{Corr}= 1$, $R>R_{\rm det}$ and $p_*>0$. To derive a 
 noncontextuality inequality, therefore, one must simply determine the precise trade-off relation satisfied by $\textrm{Corr}, R,$ and $p_*$ in a universally noncontextual model.

Applying the assumption of preparation noncontextuality to Eq.~\eqref{opeqSs2}, we can infer that
\begin{align}
\nu(\lambda) &= \mu(\lambda|\mathcal{S}_*)= \sum_{\mathcal{s}_*} \mu(\lambda| \mathcal{s}_*,\mathcal{S}_*) {\rm pr}(\mathcal{s}_*|\mathcal{S}_*).
\end{align}
Substituting this into Eq.~\eqref{ExpForCorr}, we obtain
\begin{align}
{\rm Corr} 
= 
 \sum_{ \mathcal{s}_*} {\rm pr}(\mathcal{s}_*|\mathcal{S}_*) {\rm Corr}({\mathcal{s}_*}),
\end{align}
where
\begin{align}\label{Corrsstar}
{\rm Corr}({\mathcal{s}_*}) &\equiv \sum_{\lambda} \textrm{Corr}(\lambda)  \mu(\lambda| \mathcal{s}_*,\mathcal{S}_*).
\end{align}
${\rm Corr}({\mathcal{s}_*})$ quantifies the average degree of correlation for the pairings $\{(\mathcal{S}_i ,\mathcal{M}_i) \}$ predicted by ontic state $\lambda$, averaged over the ontic states in the support of $\mu(\cdot| \mathcal{s}_*,\mathcal{S}_*)$.

The argument proceeds by showing that the quantity ${\rm Corr}({\mathcal{s}_*}=0)$ has a nontrivial upper bound.  

Define
\begin{align}
R(\lambda) = F(\{ \xi(\vec{\mathcal{m}}^{(\alpha)}| \frak{M}^{(\alpha)}, \lambda )\}_{\alpha}),
\end{align}
where $F$ is the linear function specified in Eq.~\eqref{defnR}, so that the expression for $R$ in the ontological model is
\begin{align}\label{RintermsofRlambda}
R=   \sum_{\lambda}  R(\lambda) \mu(\lambda| \mathcal{s}_*,\mathcal{S}_*)
\end{align}
Recalling that $R_{\rm det}$ denotes the maximum value that can be achieved by deterministic noncontextual assignments to the measurements, if $R>R_{\rm det}$, then some of the ontic states in the support of $\mu(\cdot| \mathcal{s}_*=0,\mathcal{S}_*)$ must be
inconsistent with a convex mixture of deterministic noncontextual assignments. In this case, ${\rm Corr}({\mathcal{s}_*}=0)$ must be bounded away from 1, 
\begin{align}\label{upperboundCsstar}
{\rm Corr}({\mathcal{s}_*}=0)<1.
\end{align}

By contrast, given that the no-go result does not make any appeal to the statistics of measurements on the preparations associated to $[\mathcal{s}_*\ne 0|\mathcal{S}_*]$,
the ontic states in the support of 
$\mu(\cdot| \mathcal{s}_*\ne 0 ,\mathcal{S}_*)$ {could potentially} assign outcomes to the measurements deterministically, which in turn implies that ${\rm Corr}({\mathcal{s}_*\ne 0})$ can only be upper bounded by  its logical maximum, 
\begin{align}
{\rm Corr}({\mathcal{s}_*\ne 0}) \le 1.
\end{align}

In all, therefore, we have 
\begin{align}\label{NCI}
{\rm Corr} \le p_* {\rm Corr}({\mathcal{s}_*}=0)  + (1-p_*).
\end{align}
Given the dependence of ${\rm Corr}({\mathcal{s}_*}=0)$ on $R$, this equation specifies a tradeoff relation between ${\rm Corr}$, $R$, and $p_*$. Such a tradeoff relation constitutes
a noncontextuality inequality derived from a statistical proof of the KS theorem. 

The precise amount by which ${\rm Corr}$ is bounded away from 1 for a given value of $R$ in Eq.~\eqref{upperboundCsstar} 
depends on two quantities: (i) the maximum value of $R(\lambda)$ for any deterministic noncontextual assignment, denoted here by $R_{\rm det}$, (ii) the maximum value of $R(\lambda)$ for any indeterministic noncontextual assignment, denoted here by $R_{\rm ind}$, and (iii) the maximum value of ${\rm Corr}(\lambda)$ for any indeterministic noncontextual assignment, which (as in the case of logical proofs of the KS theorem) we denote by ${\rm Corr}_{\rm ind}$.
The values of $R_{\rm det}$, $R_{\rm ind}$, and ${\rm Corr}_{\rm ind}$ depend on the particular statistical proof of the KS theorem one is considering.   
We will show that
\begin{align}\label{fofRgeneral}
{\rm Corr}({\mathcal{s}_*}=0) \le \frac{R_{\rm ind}-R}{R_{\rm ind}-R_{\rm det}} (1- {\rm Corr}_{\rm ind}) + {\rm Corr}_{\rm ind}.
\end{align}

Substituting this into Eq.~\eqref{NCI}, we obtain the main result of this article.
\begin{theorem}\label{mainthm} In a prepare-and-measure experiment that admits of a universally noncontextual ontological model, the following tradeoff relation between ${\rm Corr}$, $R$, and $p_*$ (defined in Eqs.~\eqref{defnCorr}, \eqref{defnR}, and \eqref{pstar}, respectively), holds:
\begin{align}\label{NCIgeneral}
{\rm Corr}\leq 1-p_* ( 1 - {\rm Corr}_{\rm ind} ) \left( \frac{R-R_{\rm det}}{R_{\rm ind}-R_{\rm det}} \right).
\end{align} 
\end{theorem}
This is our noise-robust noncontextuality inequality.

Recall that by assumption, the precise form of $R$ depends on which statistical proof of the KS theorem one is considering, and that the definition of $R$ is such that $R_{\rm ind}>R_{\rm det}$.

Note that this inequality implies that if ${\rm Corr}=1$ and $p_*>0$, then $R \le R_{\rm det}$, so that our noise-robust noncontextuality inequality for a given statistical proof of the KS theorem  reduces to the KS-noncontextuality inequality that is conventionally associated to that proof
\cite{KCBS, CSW, expt1,expt2,expt3,expt4,expt5}.
 Experimentally, however, one {\em never} achieves perfect correlation, that is, one always finds ${\rm Corr}<1$,  so that our criterion for noncontextuality {\em never} reduces to a conventional KS-noncontextuality inequality in a real experiment.
 It follows that a violation of a conventional KS-noncontextuality inequality ($R \le R_{\rm det}$) in a real experiment
is insufficient to demonstrate the failure of noncontextuality.

A question that arises at this point is whether it might still be appropriate to test the inequality $R \le R_{\rm det}$ on the grounds that it tests 
the assumption of KS-noncontextuality rather than the assumption of universal noncontextuality.    Recalling from Eq.~\eqref{KSNC} that the assumption of KS-noncontextuality incorporates an assumption of outcome determinism, to adopt such a view would be to simply {\em assume} outcome determinism rather than seeking to justify it from preparation noncontextuality and perfect correlations between sources and measurements.  However, it was shown in  Ref.~\cite{Spe14} (see also Ref.~\cite{kunjwal2017}) that assuming KS-noncontextuality (hence outcome determinism) for unsharp quantum measurements leads to absurd conclusions, such as the failure of KS-noncontextuality for experiments that are completely classical (in the sense that all states and measurements are diagonal in the same basis).  Therefore, the assumption of KS-noncontextuality is only applicable to sharp, i.e., noiseless, quantum measurements, which are never achieved experimentally.  For operational theories other than quantum theory, the same argument holds:  every measurement that can be achieved in a real experiment fails to satisfy the ideal of noiselessness, and it is only for such noiseless measurements that the assumption of KS noncontextuality is justified.\footnote{Note that the inappropriateness of applying KS-noncontextuality to real experiments has been argued in detail elsewhere over the years \cite{Spe05, KrishnaSpekkensWolfe, KunjSpek, kunjwal2017, finegen, Spe14, exptlpaper} and our comments here are meant merely to highlight the precise sense in which this plays out for statistical proofs of the KS theorem.}

The noncontextuality inequality of Eq.~\eqref{NCIgeneral}, on the other hand, accommodates noisy experimental data for which ${\rm Corr}<1$. Thus, even if sources and measurements deviate from the ideal of sharpness in an experiment, one can still see a violation of our inequality.  It is in this sense that it is robust to noise.

One can also deduce the precise limit to noise tolerance for such an inequality.  For a fixed $p_*$, 
in order to obtain a nontrivial upper bound on $R$ (i.e., a bound smaller than $R_{\rm ind}$), one must have ${\rm Corr} > 1- p_* (1- {\rm Corr}_{\rm ind})$.
If the noise is such that ${\rm Corr}$ is reduced to a value below this bound, then it becomes 
impossible to witness contextuality via this inequality. Further, if in addition to fixing $p_*$, one achieves a certain value of $R$, say $R=R_{\rm expt}$, then contextuality is witnessed if and only if ${\rm Corr} > 1- p_* (1- {\rm Corr}_{\rm ind})\left(\frac{R_{\rm expt}-R_{\rm det}}{R_{\rm ind}-R_{\rm det}}\right)$ (which is just a rewriting of the violation of Eq.~\eqref{NCIgeneral}).  
In Appendix C, we provide further details about what an experiment must achieve in order to test an inequality of the form of Eq.~\eqref{NCIgeneral}.

The proof of Eq.~\eqref{fofRgeneral} proceeds as follows. Without any loss of generality, we identify the set of ontic states $\Lambda$ with the set of vertices of the polytope of noncontextual assignments to the elements of $\frak{M}$. We divide the vertices into two sets, corresponding to deterministic and indeterministic assignments, denoted $\Lambda_{\rm det}$ and $\Lambda_{\rm ind}$ respectively, so that $\Lambda = \Lambda_{\rm det} \cup \Lambda_{\rm ind}$.  
For all $\lambda \in \Lambda_{\rm det}$, $R(\lambda)$ satisfies the nontrivial upper bound 
$R(\lambda)\leq R_{\rm det}$ (where $R_{\rm det}<R_{\rm ind}$ because $R$ is, by construction, a function that cannot achieve the logically maximal value of $R_{\rm ind}$ for deterministic noncontextual assignments) 
 while $\textrm{Corr}(\lambda)$ can always achieve its logical maximum of 1, so that the bound is trivial, $\textrm{Corr}(\lambda) \le 1$.  
 By contrast, for all $\lambda \in \Lambda_{\rm ind}$, $\textrm{Corr}(\lambda)$ satisfies the nontrivial upper bound  $\textrm{Corr}(\lambda)\le {\rm Corr}_{\rm ind}$ 
 (where ${\rm Corr}_{\rm ind}<1$ because indeterministic noncontextual assignments necessarily imply a failure to achieve perfect source-measurement correlations),
 but because there exist $\lambda \in \Lambda_{\rm ind}$ such that $R(\lambda)$ achieves its maximum of $R_{\rm ind}$, we have only the trivial bound $R(\lambda)\leq R_{\rm ind}$. 
We now make use of these facts 
to determine the upper bound on ${\rm Corr}({\mathcal{s}_*}=0)$ for a given value of $R$.
Defining 
$$\mu_{\rm det} \equiv \sum_{\lambda\in \Lambda_{\rm det}} \mu(\lambda| \mathcal{s}_*=0,\mathcal{S}_*)$$ and 
$$\mu_{\rm ind} \equiv \sum_{\lambda\in \Lambda_{\rm ind}} \mu(\lambda| \mathcal{s}_*=0,\mathcal{S}_*),$$ so that $\mu_{\rm det} + \mu_{\rm ind}= 1$, 
and recalling Eqs.~\eqref{Corrsstar} and \eqref{RintermsofRlambda}, we have 
\begin{align}
&{\rm Corr}({\mathcal{s}_*}=0) \leq \mu_{\rm det} + 
{\rm Corr}_{\rm ind} \mu_{\rm ind},\label{Corrbound}
\end{align}
 and 
 \begin{align}
 R\leq R_{\rm det} 
\mu_{\rm det} + R_{\rm ind}\mu_{\rm ind}.
\end{align}
 Eliminating $\mu_{\rm det}$ and $\mu_{\rm ind}$ from these constraints, we obtain Eq.~\eqref{fofRgeneral}.

 The noncontextuality inequality of Eq.~\eqref{NCIgeneral} can be saturated by a noncontextual ontological model in certain circumstances.\footnote{See Section VI.B of Ref.\cite{kunjwal2017} for a detailed discussion of this noncontextual ontological model.}  Define $\Lambda^{\rm max}_{\rm det} \equiv \{ \lambda \in \Lambda_{\rm det}: R(\lambda)=R_{\rm det}, {\rm Corr}(\lambda) = 1\}$ and $\Lambda^{\rm max}_{\rm ind} \equiv \{ \lambda \in \Lambda_{\rm ind}: R(\lambda)=R_{\rm ind}, {\rm Corr}(\lambda) = {\rm Corr}_{\rm ind}\}$.  Clearly, $\Lambda^{\rm max}_{\rm det} \subseteq \Lambda_{\rm det}$ and $\Lambda^{\rm max}_{\rm ind} \subseteq \Lambda_{\rm ind}$.   For noncontextual measurement-assignment polytopes based on statistical proofs of the KS theorem, $\Lambda^{\rm max}_{\rm det}$ is always a non-empty set.  If $\Lambda^{\rm max}_{\rm ind}$ is non-empty, then the noncontextuality inequality of Eq.~\eqref{NCIgeneral} can be saturated. 
The reason is that in this case one can choose the support of $\mu(\lambda| \mathcal{s}_* = 0,\mathcal{S}_*)$ on $\Lambda_{\rm det}$ to be restricted to $\Lambda^{\rm max}_{\rm det}$ and the support of $\mu(\lambda| \mathcal{s}_* = 0,\mathcal{S}_*)$ on $\Lambda_{\rm ind}$ to be restricted to $\Lambda^{\rm max}_{\rm ind}$, in which case the inequalities in Eqs.~\eqref{Corrbound} and 
\eqref{Rbound} become equalities. The condition that $\Lambda^{\rm max}_{\rm ind}$ be non-empty is satisfied for the case of odd $n$-cycle scenario, and therefore the noncontextuality inequalities that we derive in this case will be tight.  To illustrate our technique on a concrete example, we now turn to the odd $n$-cycle scenario. 

For the case of the $n$-cycle scenario,
the function $F$ defining the quantity $R$ is specified in Eq.~\eqref{defnRncycle}.  As noted in our previous discussion of the $n$-cycle scenario, for deterministic vertices of the noncontextual measurement-assignment polytope, there is a nontrivial upper bound on $R(\lambda)$, namely, $R(\lambda) \le R_{\rm det}$, where
 \begin{align}\label{Rdncycle}
R_{\rm det} = \frac{n-1}{n}.
 \end{align}
Similarly, for every indeterministic vertex of the noncontextual measurement-assignment polytope, there is a nontrivial upper bound on ${\rm Corr}(\lambda)$,   namely, ${\rm Corr}(\lambda) \le {\rm Corr}_{\rm ind}$, where
 \begin{align}\label{Corrincycle}
{\rm Corr}_{\rm ind} = \frac{1}{2}.
 \end{align}
 Also, we have 
 \begin{align}
 R_{\rm ind}=1.
 \end{align}
(For further discussion of the vertices of the polytope in the case of the $n$-cycle scenario, see Appendix~\ref{ncycleproof}.)
 Substituting Eqs.~\eqref{Rdncycle} and \eqref{Corrincycle} into Eq.~\eqref{NCIgeneral}, we find that a noise-robust noncontextuality inequality for the $n$-cycle scenario is:
\begin{align}\label{NCIncycle}
{\rm Corr}\leq 1-p_*\frac{n}{2}\left(R-\frac{n-1}{n}\right).
\end{align} 

 The quantum realization of the $n$-cycle scenario~\cite{KCBS,LSW} that was discussed earlier clearly violates this inequality.   It suffices to note that by Eq.~\eqref{NCIncycle}, if ${\rm Corr}=1$ and $p_* > 0$, then $R \le \frac{n-1}{n}$, while this quantum realization achieves ${\rm Corr}=1$, $p_*=\frac{1}{3}$  and $R= \frac{2\cos(\frac{\pi}{n})}{1+\cos(\frac{\pi}{n})}>\frac{n-1}{n}$.    The violation persists in the presence of noise: if $p_*=\frac{1}{3}$, there is a range of values of ${\rm Corr}$ and $R$ below those achieved in the ideal quantum realization --- that is,  where ${\rm Corr}<1$ and $R< \frac{2\cos(\frac{\pi}{n})}{1+\cos(\frac{\pi}{n})}$ --- such that the inequality is still violated. 
 
 \section{Discussion}
 
As noted earlier, Xu {\em et al.}~\cite{Xuetal} have previously obtained noise-robust noncontextuality inequalities starting from the KCBS~\cite{KCBS} and Yu-Oh~\cite{YuOh} statistical proofs of the KS theorem. In Appendix \ref{commentxuetal}, we compare our approach to theirs for the case of the odd $n$-cycle scenario. 
One difference is that all of the simulating measurements in their approach are 3-outcome measurements, rather than the 4-outcome case we have considered here (see also Appendix \ref{alternative}).   We also show that their inequality for the odd $n$-cycle case is a special case of our noncontextuality inequality for the odd $n$-cycle scenario, Eq.~\eqref{NCIncycle}, when $p_*$ is presumed to take the value of 1/3.  Hence, unlike Ref.~\cite{Xuetal}, our inequality does not presume that all the ensembles of preparations in the experiment correspond to uniformly random probability distributions (i.e., probability $1/3$ for each preparation).  Rather, it merely presumes that the relevant operational equivalences hold.
If the value of $p_*$ realized in an experiment
is different from the 1/3 value of the ideal quantum realization,  
then our inequality specializes 
to one that is different from that  
of Ref.~\cite{Xuetal}.
These differences are not very significant, however.  We consider the main advantage of our approach over that of Ref.~\cite{Xuetal} to be that it makes clear precisely which aspects of the polytope of noncontextual measurement-assignments (for any given statistical proof of the KS theorem) need to be identified in order to determine the form of the noncontextuality inequality.

The value of upper bounds on $R(\lambda)$ for deterministic and indeterministic vertices of the noncontextual measurement-assignment polytope, denoted here by $R_{\rm det}$ and $R_{\rm ind}$ respectively, 
 are well-studied for many statistical proofs of the KS theorem\cite{allncycles, CSW}.  
 The values of the upper bounds on ${\rm Corr}(\lambda)$ for indeterministic vertices of this polytope, denoted here by  ${\rm Corr}_{\rm ind}$, have not been studied previously, but are just as easy to determine.\footnote{Indeed, this quantity 
 makes appearance as a hypergraph invariant in Ref.~\cite{kunjwal2017}, in addition to the usual invariants in the graph-theoretic approach of Ref.~\cite{CSW}.} From these, one can determine a noncontextuality inequality for any statistical proof of the KS theorem via Eq.~\eqref{NCIgeneral}.  
A study of how our technique allows one to convert the graph-theoretic framework of \cite{CSW} to a hypergraph-theoretic framework for noise-robust noncontextuality inequalities is carried out in \cite{kunjwal2017}.

\section*{Acknowledgments} 
We would like to thank David Schmid and Elie Wolfe for discussions, and Debashis Saha and Zhen-Peng Xu for comments on an earlier version of this paper. Research at Perimeter Institute is supported by the Government of Canada through the Department of Innovation, Science and Economic Development Canada and by the Province of Ontario through the Ministry of Research, Innovation and Science.

\begin{appendix}

\section{Elaboration of the ideas underlying the technique}
\label{roleofR}
The features of the correlations within compatible subsets of $\frak{M}$ (for the special preparation $[\mathcal{s}_*=0|\mathcal{S}_*]$) which underlie the no-go theorem of Proposition \ref{nogouniversalNC} constitute a 
witness that these correlations {\em cannot} arise from a distribution $\mu(\lambda|\mathcal{s}_*=0,\mathcal{S}_*)$ supported {\em only} on ontic states  corresponding to measurement-noncontextual and outcome-deterministic assignments. 
These features are represented by a quantity $R$
that is upper bounded by a constant number if the correlations {\em do} arise from a 
distribution $\mu(\lambda|\mathcal{s}_*=0,\mathcal{S}_*)$ supported only on such ontic states.

This is precisely analogous to how, in a Bell scenario, the violation of a Bell inequality witnesses the fact that the distribution over ontic states has support on 
 indeterministic vertices of the no-signalling polytope, e.g., PR-boxes \cite{prpaper} in the case of the CHSH scenario. 
The ontic states corresponding to deterministic vertices of the no-signalling polytope define the Bell polytope. 
A Bell violation thus rules out outcome-deterministic locally-causal ontological models.  

By Fine's theorem \cite{finetheorem}, a Bell violation also rules out locally-causal ontological models that are outcome-{\em in}deterministic. This is because the notion of local causality implies factorizability of the joint response function, so that these joint response functions are convex mixtures of outcome-deterministic assignments to the local measurements. 
In other words, Fine's theorem shows that there is no loss of generality in assuming outcome determinism in tests of locality.\footnote{See Ref.~\cite{finegen} for an analysis of the role of Fine's theorem in tests of locality 
vis-\`a-vis tests of noncontextuality.} However, for tests of noncontextuality involving nontrivial applications of the assumption of measurement noncontextuality (as is the case for every test of noncontextuality arising from a statistical or logical proof of the KS theorem), there is no analogue of Fine's theorem.  The reasons for this are described in Ref.~\cite{Spe14}.

Now, for ${\rm Corr}$ to be bounded away from $1$, a non-empty subset of the ontic states in the union of the supports of sources in $\frak{S}$ must correspond to the indeterministic vertices of the polytope of measurement noncontextual assignments of probabilities to measurement outcomes. In the case of noncontextuality inequalities inspired by logical
proofs of the KS theorem \cite{KunjSpek}, {\em all} the ontic states in the union of the supports of sources in $\frak{S}$ correspond to indeterministic vertices of this polytope, simply because the polytope admits {\em no} deterministic vertices on account of the KS-uncolourability that such proofs hinge upon. On the other hand, for noncontextuality inequalities inspired by statistical proofs of the KS theorem,
  we need a witness for the fact that some non-empty subset of the union of the ontic supports of sources in $\frak{S}$ corresponds to indeterministic vertices of the measurement noncontextuality polytope. Only then can we expect ${\rm Corr}$ to be bounded away from $1$. This witness corresponds to the quantity $R$ exceeding its KS-noncontextual bound $R_{\rm det}$.
  
\section{Noncontextual measurement-assignment polytope for the $n$-cycle scenario for odd $n$}\label{ncycleproof}

\begin{figure}
	\includegraphics[scale=0.3]{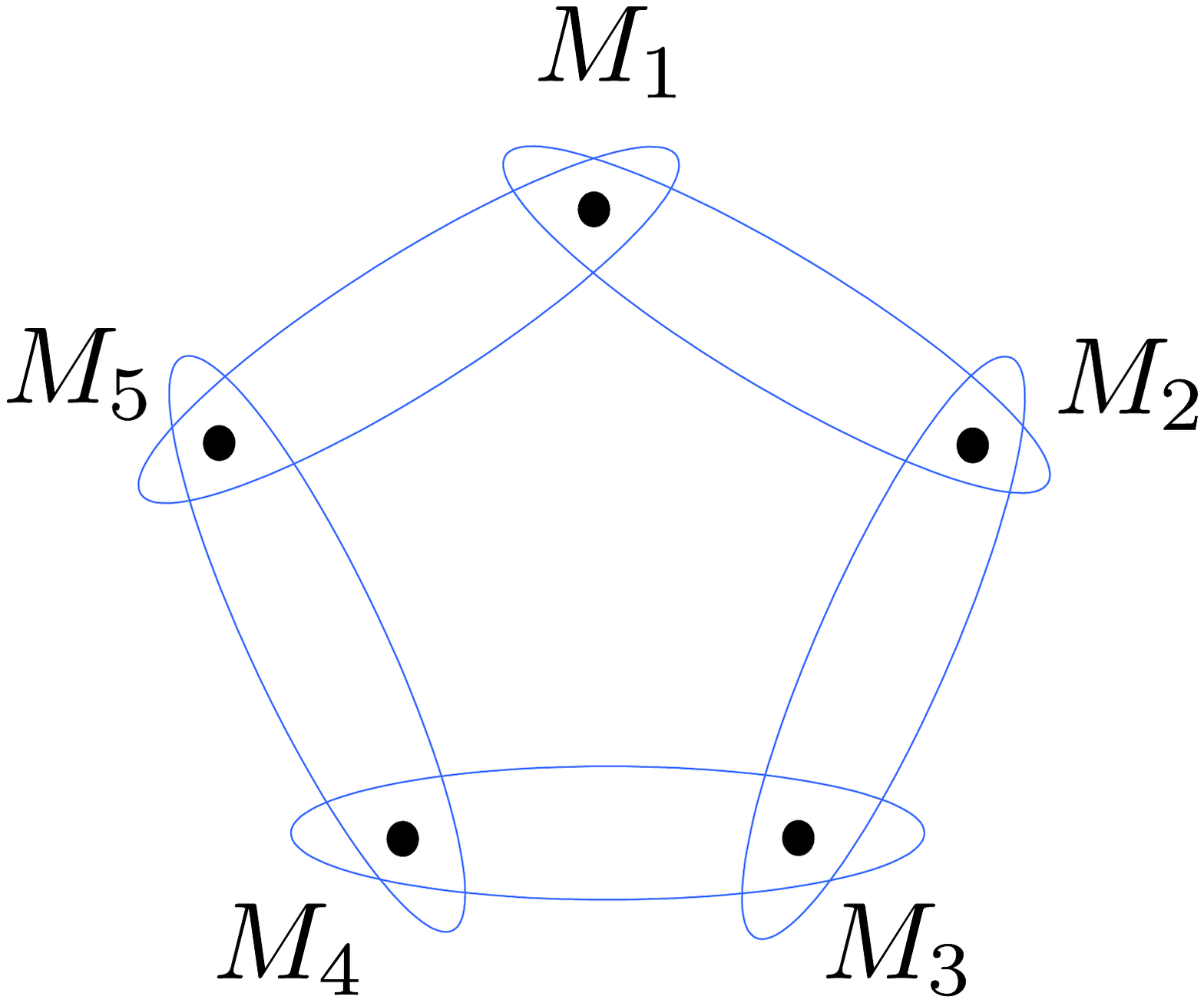}
	\caption{Compatibility relations for the 5-cycle case: the vertices represent the equivalence classes of binary-outcome measurements and the edges denote compatibility (i.e., joint measurability) of the vertices they contain.  Specifically, the $i$th vertex denotes the equivalence class $\mathcal{M}_i$, and the $i$th edge denotes the equivalence class $\mathcal{M}^{(i,i\oplus 1)}$.
	}
	\label{compatibilityhypergraph}
\end{figure}

An ontological model must specify a conditional probability distribution for every compatible subset of measurements.  In the $n$-cycle scenario, there are $n$ such subsets, corresponding to all adjacent pairs of measurements in the cycle.   This is depicted in Fig.~\ref{compatibilityhypergraph}. Recalling that $\mathcal{M}^{(i,i \oplus 1)}$ denotes the equivalence class of measurement procedures that jointly simulate $\mathcal{M}_{i}$ and $\mathcal{M}_{i \oplus 1}$, an ontological model must specify an $n$-tuple of response functions of the form
\begin{align}
\xi(m_i, m_{i \oplus 1}| \mathcal{M}^{(i,i \oplus 1)}, \lambda).
\end{align}
Marginalization of an outcome of a procedure is modelled in the ontological model by marginalization of the response function, so that the ontological representation of $\mathcal{M}_i$ satisfies
\begin{align}
\xi(m_i | \mathcal{M}_i, \lambda) = 
\sum_{m_{i \oplus 1}} \xi(m_i, m_{i \oplus 1}| \mathcal{M}^{(i,i \oplus 1)}, \lambda).\label{margrf}
\end{align}
The assumption of noncontextuality implies that the response function representing a measurement depends only on its equivalence class, so that we infer the constraints 
\begin{align}
\forall i : &\sum_{m_{i \ominus 1}} \xi(m_i, m_{i \ominus 1}| \mathcal{M}^{(i,i \ominus 1)}, \lambda) \nonumber\\
&= 
\sum_{m_{i \oplus 1}} \xi(m_i, m_{i \oplus 1}| \mathcal{M}^{(i,i \oplus 1)}, \lambda).\label{const2}
\end{align}
Together with the conditions of being a probability distribution, $0\le  \xi(m_i, m_{i \oplus 1}| \mathcal{M}^{(i,i \oplus 1)}, \lambda) \le1$, and of normalization, $\sum_{m_i,m_{i \oplus 1}} \xi(m_i, m_{i \oplus 1}| \mathcal{M}^{(i,i \oplus 1)}, \lambda) =1$, Eq.~\eqref{const2} defines the constraints on the response functions.  The set of solutions to these constraints  in turn determines the set of solutions for the $n$-tuple of response functions for the binary-outcome measurements, $\{ \xi( \mathcal{m}_i | \mathcal{M}_i, \lambda) \}_i$, through Eq~\eqref{margrf}. 

For every $\lambda$, such an $n$-tuple of response functions defines a possible $n$-tuple of (deterministic or indeterministic) assignments to all of the measurements.   Following Ref.~\cite{KrishnaSpekkensWolfe}, we term the latter set the noncontextual measurement-assignment polytope. (It is equivalent to what is termed the ``no-disturbance'' polytope elsewhere; for the $n$-cycle scenario, it was characterized in Ref.~\cite{allncycles}.)

There are two types of vertex for this polytope, corresponding to noncontextual measurement assignments that are deterministic and indeterministic respectively.  If we identify the set of ontic states $\Lambda$ with the set of vertices of the polytope, as in the main text, then the two types define a partition of $\Lambda$ into subsets $\Lambda_{\rm det}$ and $\Lambda_{\rm ind}$. 

\begin{figure}
	\includegraphics[scale=0.3]{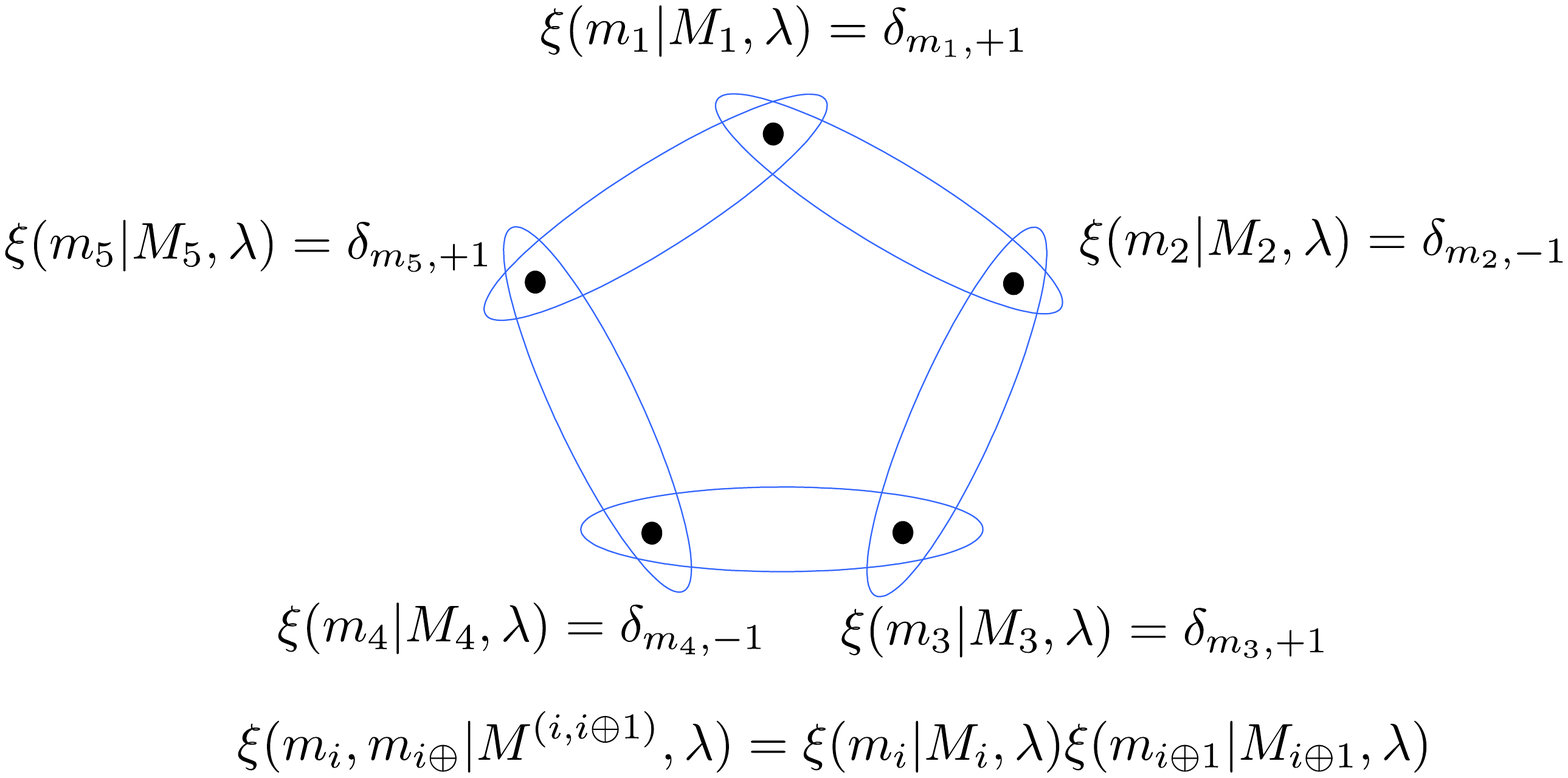}
	\caption{An example of a deterministic measurement-noncontextual assignment that maximizes the amount of anticorrelation achievable by such assignments, $R(\lambda)= \frac{4}{5}$.}\label{kappa1}
\end{figure}

The deterministic vertices are simply those that assign an outcome to each of the $\mathcal{M}_i$ independently.  There are consequently $2^n$ of these.  All of these  can achieve the logical maximum value of 1 for ${\rm Corr}(\lambda)$,  
and of these, there are $2n$ that achieve $R(\lambda)= \frac{n-1}{n}$.  For instance, this is achieved for the vertices $\{\kappa_j\}_{j=1}^{n}$ of the form
\begin{align}
\forall i\in [n] : &\xi( m_i , m_{i\oplus 1} |\kappa_j)  \nonumber\\
&= \xi( \mathcal{m}_i |\kappa_j) \xi( \mathcal{m}_{i\oplus 1} |\kappa_j) 
\end{align}
where 
\begin{align}
\xi( \mathcal{m}_i |\kappa_j) 
&= \delta_{\mathcal{m}_i,+1} \; {\rm for}\; i \in \{ j, j\oplus 2, \dots, j\oplus (n-1)\} \nonumber\\
&= \delta_{\mathcal{m}_i,-1} \; {\rm for}\; i \in \{ j\oplus 1, j\oplus 3, \dots, j\oplus (n-2)\}.
\end{align}
 $\kappa_1$ is depicted in Fig.~\ref{kappa1}.

To obtain another set of $n$ vertices that achieve $R(\lambda)= \frac{n-1}{n}$, it suffices to flip the sign of all the assignments.  

These $2n$ vertices constitute the  subset of $\Lambda_{\rm det}$ that can saturate the inequalities ${\rm Corr}(\lambda) \le 1$ and $R(\lambda) \le \frac{n-1}{n}$ described in the main text.

\begin{figure}
	\includegraphics[scale=0.3]{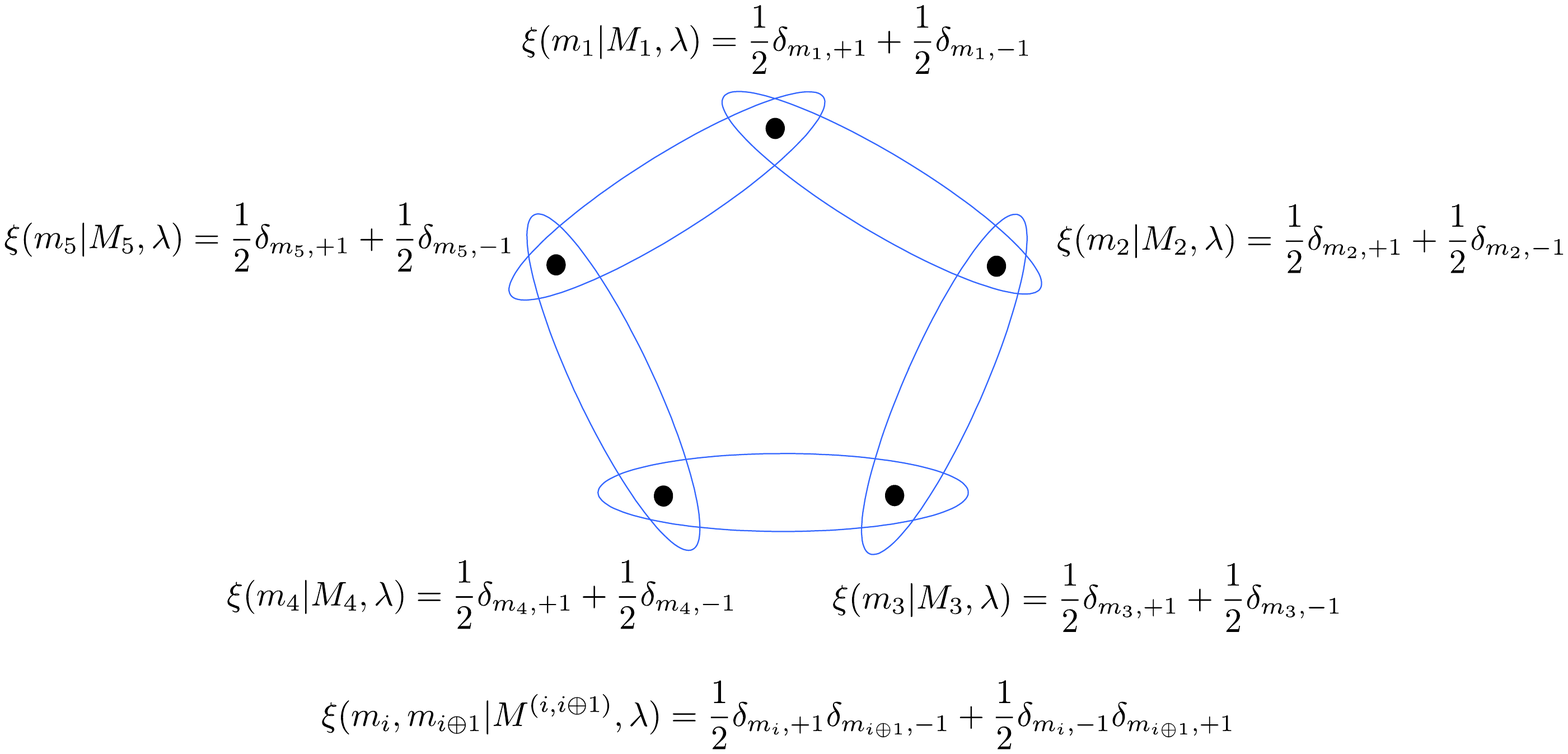}
	\caption{The indeterministic measurement-noncontextual assignment that achieves perfect anticorrelation, $R(\lambda)= 1$.}
	\label{kappastar}
\end{figure}

The indeterministic vertices are those that exhibit either  perfect positive correlation or perfect negative correlation for each of the adjacent pairs of measurements with the number of pairs that exhibit perfect negative correlation being odd.  There are $2^{n-1}$ such assignments.  All of these achieve ${\rm Corr}(\lambda) = \frac{1}{2}$.  One of them also achieves $R(\lambda)=1$, namely, the one, denoted $\kappa_*$, corresponding to perfect negative correlation for all pairs,
\begin{align}
\forall i\in [n] : &\xi( m_i, m_{i \oplus} |\kappa_*)  \nonumber\\
&=  \frac{1}{2} \delta_{m_i,+1} \delta_{m_{i\oplus},-1}
+ \frac{1}{2} \delta_{m_i,-1} \delta_{m_{i\oplus},+1},
\end{align}
such that the assignment to each $\mathcal{M}_i$ is uniformly random,
\begin{align}
\forall i\in [n] : &\xi( \mathcal{m}_i |\kappa_*) =  \frac{1}{2} \delta_{\mathcal{m}_i,-1} +\frac{1}{2} \delta_{\mathcal{m}_i,+1}.
\end{align}
 The vertex $\kappa_*$ is depicted in Fig.~\ref{kappastar}.   It is the only element of the set $\Lambda_{\rm ind}$ that saturates the inequalities ${\rm Corr}(\lambda) \le \frac{1}{2}$ and $R(\lambda) \le 1$ described in the main text. 

For each statistical proof of the KS theorem, one can determine the polytope of noncontextual measurement assignments.  The details of this polytope will determine the nontrivial upper bound on $R(\lambda)$ for the deterministic vertices (where $R$ is defined in a manner that is specific to the statistical proof one is considering; see Eq.~\eqref{defnR} for the general form and Eq.~\eqref{defnRncycle} for an example from the $n$-cycle scenario) and the nontrivial upper bound on ${\rm Corr}(\lambda)$ for the indeterministic vertices.  These determinations are all that  one requires to derive a noncontextuality inequality for any given statistical proof of the KS theorem. 

\section{How to test such inequalities experimentally}\label{exptappendix}

Recall that $\alpha$ is a variable that runs over the compatible subsets of measurements in $\frak{M}$ and that $\mathcal{M}^{(\alpha)}$ denotes the measurement that jointly simulates the compatible subset associated to $\alpha$, that is, $\{ \mathcal{M}_{i} \}_{i\in \alpha}$.  Strictly speaking, $\mathcal{M}^{(\alpha)}$ denotes an equivalence class of measurement procedures. Let $M^{(\alpha)}$ denote a specific procedure in the class $\mathcal{M}^{(\alpha)}$, and let $M_{i(\alpha)}$ denote the procedure in the equivalence class $\mathcal{M}_i$ that is obtained by implementing the joint measurement procedure $M^{(\alpha)}$ and post-processing its outcome (specifically, by marginalizing over the outcomes of all measurements other than $\mathcal{M}_i$ in the compatible subset).   Supposing that $\alpha=a_i$ and $\alpha=a'_i$ both correspond to compatible subsets of measurements that include $\mathcal{M}_i$, then $M_{i(\alpha=a_i)}$ and $M_{i(\alpha=a'_i)}$ are distinct procedures in the operational equivalence class $\mathcal{M}_i$.

 Any  experiment that involves the set of measurements $\frak{M}$ and seeks to test noncontextuality must aim to implement a specific measurement procedure $M^{(\alpha)}$ for each $\alpha$ such that every operational equivalence relation of the form $M_{i(\alpha=a_i)} \simeq M_{i(\alpha=a'_i)}$ holds. 
 
For instance, in the case of the $n$-cycle scenario, where $\mathcal{M}^{(i,i \oplus 1)}$ denotes the equivalence class of measurement procedures that jointly simulate $\mathcal{M}_{i}$ and $\mathcal{M}_{i \oplus 1}$, where $M^{(i,i \oplus 1)}$ denotes a specific procedure in the class $\mathcal{M}^{(i,i \oplus 1)}$, and where $M_{i(i\oplus 1)}$ denotes the procedure one obtains by implementing $M^{(i,i\oplus 1)}$ and marginalizing over the outcome $m_{i\oplus 1}$,
 any  experiment that seeks to test noncontextuality 
  must aim to implement a specific measurement procedure $M^{(i,i\oplus 1)}$ for each $i\in [n]$ such that the operational equivalence relation $M_{i(i\oplus 1)} \simeq M_{i(i\ominus 1)}$ holds for all $i\in [n]$.  
 
Furthermore,  any  experiment that involves the set of sources $\frak{S}$ and seeks to test noncontextuality must aim to implement a specific binary-outcome source procedure $S_i$  for each $i $, as well as a special source procedure $S_*$ such that the operational equivalence relations $[\top | S_i] \simeq [\top | S_{i'}] \simeq [ \top |S_*]$ for all $i,i'$ hold (see Eq.~\eqref{opeqSs2}).

Whichever measurement procedures $\{ M^{(\alpha)}\}_{\alpha}$ and source procedures $\{S_i\}_i$ and $S_*$ one targets, however, experimental imperfections ensure that the relevant operational equivalence relations are not achieved precisely.   
 But given that noncontextuality is an inference from operational equivalences to equivalences in the ontological model, such imprecision blocks the derivation of any consequences for the ontological model of the experiment. This was termed the problem of no strict operational equivalences in Ref.~\cite{exptlpaper}.  It was shown there how to solve it using the technique of secondary procedures (see also Sec.~V of Ref.~\cite{KrishnaSpekkensWolfe}).  The idea is to identify, within the convex hull of the sources and measurements that were experimentally implemented (termed the primary procedures),  sources and measurements that satisfy the operational equivalence relations {\em exactly} (termed the secondary procedures), and then to test the noncontextuality inequalities on the secondary procedures.  
 
Note that operational equivalence of two measurements (sources) requires equivalence of statistics for {\em all} sources (measurements), or equivalently, equivalence for a tomographically complete set of sources (measurements).  Experiments seeking to test operational equivalence relations, therefore, must accumulate evidence in favour of a given set of procedures being tomographically complete.  See, e.g., the evidence described in Ref.~\cite{exptlpaper}.  Note that there is a loophole, which we term the {\em tomography loophole}, for experiments testing universal noncontextuality: no matter how much evidence one accumulates for the tomographic completeness of some set of procedures, it is possible that future experiments will uncover new procedures whose statistics are {\em not} predicted by the statistics of those in the set.  It follows that any hypothesis of tomographic completeness of some set is necessarily tentative.  One should endeavour to falsify it experimentally, and as long as it resists falsification, one has good evidence for the hypothesis.  But  one can never verify it.

\section{An alternative way of operationalizing the KCBS proof of the KS theorem}
\label{alternative}

In this article, we have operationalized the KCBS proof of the KS theorem as an $n$-cycle scenario with odd $n$, that is, as $n$ binary-outcome measurements arranged in a cycle such that adjacent pairs are jointly measurable.  The outcome set of each joint measurement can be taken to be the Cartesian product of the outcome sets of the two measurements being simulated, so that each joint measurement has four outcomes.  Recall that the binary-outcome measurements were denoted $\mathcal{M}_i$ with outcome $\mathcal{m}_i$, and the joint measurements were denoted  $\mathcal{M}^{(i,i\oplus 1)}$ with outcome $(\mathcal{m}_i,\mathcal{m}_{i\oplus 1})$ (For more details, see Appendix \ref{ncycleproof}.)

However, one can also imagine operationalizing the proof as an odd number $n$ of three-outcome measurements.  Denoting the $i$th measurement by $\mathcal{M}^{\rm tri}_i$, and taking the outcome set to be $\mathcal{m}^{\rm tri}_i\in \{0,1,2\}$, the operational equivalence relations have the form $\forall i : [0|M^{\rm tri}_i] \simeq [2|M^{\rm tri}_{i \oplus 1}]$.  In words, the last outcome of one measurement in the cycle is operationally equivalent to the first outcome of the next measurement in the cycle. 

To translate between the two approaches, it suffices to recognize that the three-outcome measurement $\mathcal{M}^{\rm tri}_i$ can be identified with the four-outcome measurement $\mathcal{M}^{(i,i\oplus1)}$ as long as one of the outcomes of the latter has probability zero for all preparations.  Specifically, we take $(\mathcal{m}_i=+1,\mathcal{m}_{i\oplus1}=+1)$ to be the outcome that is always assigned zero probability, and make the translations $m^{\rm tri}_i=0 \leftrightarrow (\mathcal{m}_i=+1,\mathcal{m}_{i\oplus1}=-1)$, $m^{\rm tri}_i=1 \leftrightarrow (\mathcal{m}_i=-1,\mathcal{m}_{i\oplus1}=-1)$, and $m^{\rm tri}_i=2 \leftrightarrow (\mathcal{m}_i=-1,\mathcal{m}_{i\oplus1}=+1)$.

If an experiment implements $n$ four-outcome measurements where all of the outcomes have nonzero probability, then one must use the $n$-cycle approach, whereas if the measurements that are implemented have only three outcomes that ever occur, then either approach can be used.

Some illustrative quantum examples help to clarify these ideas.  

Recall that in KCBS's quantum construction~\cite{KCBS},
 the binary-outcome measurement $\mathcal{M}_i$ is represented by a projection-valued measure (PVM) $\{ \Pi^{(i)}_+ , \Pi^{(i)}_- \}$ where $\Pi^{(i)}_+ \equiv|l_i\rangle\langle l_i|$, and $\Pi^{(i)}_- = I- \Pi^{(i)}_+$. By construction, any two neighbouring PVMs, $\{ \Pi^{(i)}_+ , \Pi^{(i)}_- \}$ and $\{ \Pi^{(i\oplus 1)}_+ , \Pi^{(i\oplus 1)}_- \}$, consist of projectors that commute and satisfy 
 $\Pi^{(i)}_+ \Pi^{(i\oplus1)}_+=0$.  It follows that the joint measurement $\mathcal{M}^{(i,i\oplus1)}$ is represented 
 (uniquely) by the four-outcome PVM consisting of the products of the projectors from each of the neighbouring PVMs, $\{ \Pi^{(i)}_+ \Pi^{(i\oplus 1)}_+ ,  \Pi^{(i)}_+ \Pi^{(i\oplus 1)}_- , \Pi^{(i)}_- \Pi^{(i\oplus 1)}_+, \Pi^{(i)}_- \Pi^{(i\oplus 1)}_-\}$.  But this simplifies to $\{ 0 ,  \Pi^{(i)}_+ , \Pi^{(i\oplus 1)}_+, I -  \Pi^{(i)}_+ - \Pi^{(i\oplus 1)}_+\}$, so that it is clear that the first outcome always has probability zero of occuring and consequently the joint measurement is translatable into one of the three-outcome variety.   This, therefore, is an example of the type described above, wherein the outcome $(\mathcal{m}_i=+1,\mathcal{m}_{i\oplus1}=+1)$ of the four-outcome joint measurement $\mathcal{M}^{(i,i\oplus1)}$  never occurs.

Now consider quantum realizations of the $n$-cycle scenario wherein the binary-outcome measurement $\mathcal{M}_i$ is {\em not} represented projectively, but rather by a nonprojective POVM, which we denote by $\{ E^{(i)}_+ , E^{(i)}_- \}$, where $E^{(i)}_- = I - E^{(i)}_+$.  The constraint that $\mathcal{M}_i$ and $\mathcal{M}_{i\oplus 1}$ be jointly simulatable implies that there must exist a four-outcome POVM, $\{G^{(i)}_{++},G^{(i)}_{+-},G^{(i)}_{-+},G^{(i)}_{--}\}$ such that $G^{(i)}_{++}+G^{(i)}_{+-}=E^{(i)}_+$ and $G^{(i)}_{++}+G^{(i)}_{-+}=E^{(i\oplus1)}_+$.  

For certain compatible pairs of nonprojective POVMs, namely, those for which $E^{(i)}_+ + E^{(i\oplus1)}_+
\leq I$, there exists a joint measurement POVM of the form $\{G^{(i)}_{++},G^{(i)}_{+-},G^{(i)}_{-+},G^{(i)}_{--}\}$ where $G^{(i)}_{++}=0$, $G^{(i)}_{+-}=E^{(i)}_+$, $G^{(i)}_{-+}=E^{(i\oplus1)}_+$, and $G^{(i)}_{--}=I-E^{(i)}_+ - E^{(i\oplus 1)}_+$.  This constitutes another example of a four-outcome joint measurement where the first outcome never occurs, so that it is translatable into one of the three-outcome variety. 

On the other hand, for {\em generic} compatible pairs of nonprojective POVMs, i.e., those for which it is not the case that $E^{(i)}_+ + E^{(i\oplus1)}_+ \leq I$, the joint measurement must be represented by a genuinely four-outcome POVM.    It follows that if one considers a quantum realization of the $n$-cycle  wherein the compatible pairs are of this sort, then the four-outcome joint measurement is not translatable into one of the three-outcome variety.

\section{Comparison with the approach of Xu {\em et. al.}}
\label{commentxuetal}

We here compare our approach to obtaining inequalities for universal noncontextuality from statistical proofs of the KS theorem to the one described in Xu et.~al.~\cite{Xuetal}.  

In fact, Ref.~\cite{Xuetal} describes {\em two} approaches to doing so.  The first is described in Sections IV A and B of that paper  and the second is described in Section V.
Neither approach, however, is presented in a manner that fully excises reference to the ideal quantum realization. 

 The first technique, for instance, makes explicit reference to predictions of quantum theory when determining the upper bound on their quantity $\mathcal{A}$ (in the case of the KCBS proof); they presume that their quantity $\mathcal{I}$ can achieve the maximum quantum value of $\sqrt{5}$ (their Eq. (17)).  
 
In the second technique (also in the case of the $n$-cycle proof), the choice of coefficients in the inequality is particular to the case when the source outcomes are uniformly random, as in the ideal quantum realization.\footnote{Note that some of the commentary provided in Ref.~\cite{Xuetal} on their derivation of the inequality may create the impression that it is important that two preparation procedures ($P$ and $\bar{P}$ in their notation) are perfectly distinguishable, so that their ontic supports are disjoint.  See, e.g., the comment above Eq.~(20) in their article.  This idealization is not, however, required to derive the inequality.}

We are confident that generalizations of the two techniques can address these concerns, because the intuition behind them (articulated in the last paragraph of Sec. 2.2 of Ref.~\cite{Xuetal}) is in line with that of our approach. However, Xu {\em et al.} did not disentangle this intuition from features of the ideal quantum realization as cleanly as we do here. 

We consider the main advantages of the approach described in this article, relative to those of Ref~\cite{Xuetal}, to be two-fold: (i) we have derived the inequalities in a principled manner, motivating the logic with a no-go theorem for universal noncontextuality, and excising all features of the ideal quantum realization that are not needed to derive nontrivial inequalities (such as the particular choice of probabilities for source outcomes),
and (ii) we have described explicitly which parameters in the noncontextuality inequality depend on the choice of statistical proof and how to compute these parameters by characterizing the noncontextual measurement-assignment polytope associated to that proof.   

For the case of the KS theorem based on the odd $n$-cycle scenario, the second technique described in Xu {\em et al.} leads to an inequality that is a special case of our noncontextuality inequality for this scenario when some parameters are fixed.  In the rest of this section, we make the connection explicit.

First, we note that the manner in which the KCBS statistical proof of the KS theorem is operationalized in Ref.~\cite{Xuetal} differs from the manner in which we do so here in precisely the sort of way outlined in the previous Appendix.  Strictly speaking, therefore, their inequality is only applicable for experiments that aim to implement a set of $n$ three-outcome measurements, $\{ \mathcal{M}^{\rm tri}_i \}_i$, with operational equivalence relations of the form $\forall i : [0|M^{\rm tri}_i] \simeq [2|M^{\rm tri}_{i \oplus 1}]$.  

For ease of comparison with our results, however, we conceptualize these three-outcome measurements as four-outcome measurements wherein one of the outcomes never occurs, and we make the particular identification between outcomes outlined in the previous Appendix.  We can then rewrite their inequalities using the notational conventions of this article.

Doing so, the inequality of Ref.~\cite{Xuetal} becomes
\begin{equation}\label{NCIXuetal}
{\rm Corr}\leq 1-\frac{n}{6}\left(R-\frac{n-1}{n}\right).
\end{equation} 
which is a special case of our inequality for the $n$-cycle scenario (Eq.~\eqref{NCIncycle}) where $p_*$  is presumed to take the value that it takes in the ideal quantum realization of the no-go result for universal noncontextuality, namely $p_*=\frac{1}{3}$.

\end{appendix}

\end{document}